\documentclass[aps,prl,twocolumn,superscriptaddress,preprintnumbers,amsmath,amssymb]{revtex4}

\usepackage{graphicx} 
\usepackage{dcolumn}  
\usepackage{rotating}
\usepackage{booktabs}
\usepackage{SIunits} 
\usepackage{graphicx}
\usepackage{dcolumn}
\usepackage{diagbox}
\usepackage{bm}
\usepackage{color}
\usepackage[english]{babel}
\usepackage{titlesec}
\usepackage{multirow}
\usepackage[colorlinks,linkcolor=blue,urlcolor=blue,anchorcolor=green,citecolor=blue,breaklinks=true]{hyperref}
\usepackage{epstopdf}
\usepackage{enumitem}
\usepackage{hhline}
\usepackage{array}
\usepackage{threeparttable}

\newcommand{\AW}{{\cal A}}
\newcommand{\pip}{\pi^+}

\newcommand{\EE}{e^+e^-}
\begin{document}
\hyphenpenalty=10000
\tolerance=1000
\vspace*{-3\baselineskip}



\title{
\boldmath Measurements of the branching fractions of the semileptonic decays $\Xi_{c}^{0} \to \Xi^{-} \ell^{+} \nu_{\ell}$ and the asymmetry parameter of $\Xi_{c}^{0} \to \Xi^{-} \pi^{+}$}



\noaffiliation
\affiliation{Department of Physics, University of the Basque Country UPV/EHU, 48080 Bilbao}
\affiliation{University of Bonn, 53115 Bonn}
\affiliation{Brookhaven National Laboratory, Upton, New York 11973}
\affiliation{Budker Institute of Nuclear Physics SB RAS, Novosibirsk 630090}
\affiliation{Faculty of Mathematics and Physics, Charles University, 121 16 Prague}
\affiliation{Chonnam National University, Gwangju 61186}
\affiliation{University of Cincinnati, Cincinnati, Ohio 45221}
\affiliation{Deutsches Elektronen--Synchrotron, 22607 Hamburg}
\affiliation{Duke University, Durham, North Carolina 27708}
\affiliation{University of Florida, Gainesville, Florida 32611}
\affiliation{Department of Physics, Fu Jen Catholic University, Taipei 24205}
\affiliation{Key Laboratory of Nuclear Physics and Ion-beam Application (MOE) and Institute of Modern Physics, Fudan University, Shanghai 200443}
\affiliation{Justus-Liebig-Universit\"at Gie\ss{}en, 35392 Gie\ss{}en}
\affiliation{Gifu University, Gifu 501-1193}
\affiliation{SOKENDAI (The Graduate University for Advanced Studies), Hayama 240-0193}
\affiliation{Gyeongsang National University, Jinju 52828}
\affiliation{Department of Physics and Institute of Natural Sciences, Hanyang University, Seoul 04763}
\affiliation{University of Hawaii, Honolulu, Hawaii 96822}
\affiliation{High Energy Accelerator Research Organization (KEK), Tsukuba 305-0801}
\affiliation{Higher School of Economics (HSE), Moscow 101000}
\affiliation{Forschungszentrum J\"{u}lich, 52425 J\"{u}lich}
\affiliation{IKERBASQUE, Basque Foundation for Science, 48013 Bilbao}
\affiliation{Indian Institute of Science Education and Research Mohali, SAS Nagar, 140306}
\affiliation{Indian Institute of Technology Guwahati, Assam 781039}
\affiliation{Indian Institute of Technology Hyderabad, Telangana 502285}
\affiliation{Indian Institute of Technology Madras, Chennai 600036}
\affiliation{Indiana University, Bloomington, Indiana 47408}
\affiliation{Institute of High Energy Physics, Chinese Academy of Sciences, Beijing 100049}
\affiliation{Institute of High Energy Physics, Vienna 1050}
\affiliation{Institute for High Energy Physics, Protvino 142281}
\affiliation{INFN - Sezione di Napoli, 80126 Napoli}
\affiliation{INFN - Sezione di Torino, 10125 Torino}
\affiliation{Advanced Science Research Center, Japan Atomic Energy Agency, Naka 319-1195}
\affiliation{J. Stefan Institute, 1000 Ljubljana}
\affiliation{Institut f\"ur Experimentelle Teilchenphysik, Karlsruher Institut f\"ur Technologie, 76131 Karlsruhe}
\affiliation{Kavli Institute for the Physics and Mathematics of the Universe (WPI), University of Tokyo, Kashiwa 277-8583}
\affiliation{Department of Physics, Faculty of Science, King Abdulaziz University, Jeddah 21589}
\affiliation{Kitasato University, Sagamihara 252-0373}
\affiliation{Korea Institute of Science and Technology Information, Daejeon 34141}
\affiliation{Korea University, Seoul 02841}
\affiliation{Kyoto Sangyo University, Kyoto 603-8555}
\affiliation{Kyungpook National University, Daegu 41566}
\affiliation{Universit\'{e} Paris-Saclay, CNRS/IN2P3, IJCLab, 91405 Orsay}
\affiliation{P.N. Lebedev Physical Institute of the Russian Academy of Sciences, Moscow 119991}
\affiliation{Liaoning Normal University, Dalian 116029}
\affiliation{Faculty of Mathematics and Physics, University of Ljubljana, 1000 Ljubljana}
\affiliation{Ludwig Maximilians University, 80539 Munich}
\affiliation{Luther College, Decorah, Iowa 52101}
\affiliation{Malaviya National Institute of Technology Jaipur, Jaipur 302017}
\affiliation{University of Maribor, 2000 Maribor}
\affiliation{Max-Planck-Institut f\"ur Physik, 80805 M\"unchen}
\affiliation{University of Mississippi, University, Mississippi 38677}
\affiliation{Moscow Physical Engineering Institute, Moscow 115409}
\affiliation{Graduate School of Science, Nagoya University, Nagoya 464-8602}
\affiliation{Universit\`{a} di Napoli Federico II, 80126 Napoli}
\affiliation{Nara Women's University, Nara 630-8506}
\affiliation{National Central University, Chung-li 32054}
\affiliation{National United University, Miao Li 36003}
\affiliation{Department of Physics, National Taiwan University, Taipei 10617}
\affiliation{H. Niewodniczanski Institute of Nuclear Physics, Krakow 31-342}
\affiliation{Nippon Dental University, Niigata 951-8580}
\affiliation{Niigata University, Niigata 950-2181}
\affiliation{Novosibirsk State University, Novosibirsk 630090}
\affiliation{Osaka City University, Osaka 558-8585}
\affiliation{Pacific Northwest National Laboratory, Richland, Washington 99352}
\affiliation{Panjab University, Chandigarh 160014}
\affiliation{University of Pittsburgh, Pittsburgh, Pennsylvania 15260}
\affiliation{Punjab Agricultural University, Ludhiana 141004}
\affiliation{Research Center for Nuclear Physics, Osaka University, Osaka 567-0047}
\affiliation{Department of Modern Physics and State Key Laboratory of Particle Detection and Electronics, University of Science and Technology of China, Hefei 230026}
\affiliation{Seoul National University, Seoul 08826}
\affiliation{Soochow University, Suzhou 215006}
\affiliation{Soongsil University, Seoul 06978}
\affiliation{Sungkyunkwan University, Suwon 16419}
\affiliation{School of Physics, University of Sydney, New South Wales 2006}
\affiliation{Department of Physics, Faculty of Science, University of Tabuk, Tabuk 71451}
\affiliation{Tata Institute of Fundamental Research, Mumbai 400005}
\affiliation{Department of Physics, Technische Universit\"at M\"unchen, 85748 Garching}
\affiliation{School of Physics and Astronomy, Tel Aviv University, Tel Aviv 69978}
\affiliation{Toho University, Funabashi 274-8510}
\affiliation{Department of Physics, Tohoku University, Sendai 980-8578}
\affiliation{Earthquake Research Institute, University of Tokyo, Tokyo 113-0032}
\affiliation{Department of Physics, University of Tokyo, Tokyo 113-0033}
\affiliation{Tokyo Institute of Technology, Tokyo 152-8550}
\affiliation{Virginia Polytechnic Institute and State University, Blacksburg, Virginia 24061}
\affiliation{Wayne State University, Detroit, Michigan 48202}
\affiliation{Yamagata University, Yamagata 990-8560}
\affiliation{Yonsei University, Seoul 03722}

\author{Y.~B.~Li}\affiliation{Key Laboratory of Nuclear Physics and Ion-beam Application (MOE) and Institute of Modern Physics, Fudan University, Shanghai 200443} 
\author{C.~P.~Shen}\affiliation{Key Laboratory of Nuclear Physics and Ion-beam Application (MOE) and Institute of Modern Physics, Fudan University, Shanghai 200443} 
\author{I.~Adachi}\affiliation{High Energy Accelerator Research Organization (KEK), Tsukuba 305-0801}\affiliation{SOKENDAI (The Graduate University for Advanced Studies), Hayama 240-0193} 
\author{K.~Adamczyk}\affiliation{H. Niewodniczanski Institute of Nuclear Physics, Krakow 31-342} 
\author{H.~Aihara}\affiliation{Department of Physics, University of Tokyo, Tokyo 113-0033} 
\author{S.~Al~Said}\affiliation{Department of Physics, Faculty of Science, University of Tabuk, Tabuk 71451}\affiliation{Department of Physics, Faculty of Science, King Abdulaziz University, Jeddah 21589} 
\author{D.~M.~Asner}\affiliation{Brookhaven National Laboratory, Upton, New York 11973} 
\author{T.~Aushev}\affiliation{Higher School of Economics (HSE), Moscow 101000} 
\author{R.~Ayad}\affiliation{Department of Physics, Faculty of Science, University of Tabuk, Tabuk 71451} 
\author{V.~Babu}\affiliation{Deutsches Elektronen--Synchrotron, 22607 Hamburg} 
% ...
\collaboration{The Belle Collaboration}

\begin{abstract}

Using data samples of 89.5 and 711 fb$^{-1}$ recorded at energies of $\sqrt{s}=10.52$ and $10.58$ GeV, respectively, with the Belle detector at the KEKB $e^+e^-$ collider,
we report measurements of branching fractions of semileptonic decays $\Xi_{c}^{0} \to \Xi^{-} \ell^{+} \nu_{\ell}$ ($\ell=e$ or $\mu$) and the $CP$-asymmetry parameter of
$\Xi_{c}^{0} \to \Xi^{-} \pi^{+}$ decay.
The branching fractions are measured to be ${\cal B}(\Xi_{c}^{0} \to \Xi^{-} e^{+} \nu_{e})=(1.31 \pm 0.04 \pm 0.07 \pm 0.38)\%$ and ${\cal B}(\Xi_{c}^{0} \to \Xi^{-} \mu^{+} \nu_{\mu})=(1.27 \pm 0.06 \pm 0.10 \pm 0.37)\%$, and the decay parameter $\alpha_{\Xi\pi}$ is measured to be $0.63 \pm 0.03 \pm 0.01$ with much improved precision compared to the current world average. The corresponding ratio ${\cal B}(\Xi_{c}^{0} \to \Xi^{-} e^{+} \nu_{e})/{\cal B}(\Xi_{c}^{0} \to \Xi^{-} \mu^{+} \nu_{\mu})$
is $1.03 \pm 0.05\pm 0.07$, which is consistent with the expectation of lepton flavor universality. The first measured asymmetry parameter ${\cal A}_{CP} = (\alpha_{\Xi^{-}\pi^{+}} + \alpha_{\bar{\Xi}^{+}\pi^{-}})/(\alpha_{\Xi^{-}\pi^{+}} - \alpha_{\bar{\Xi}^{+}\pi^{-}}) = 0.024 \pm 0.052 \pm 0.014$ is found to be consistent with zero. The first and the second uncertainties above are statistical and systematic, respectively, while the third ones arise due to the uncertainty of the $\Xi_{c}^{0} \to \Xi^{-} \pi^+$ branching fraction.
\end{abstract}


\maketitle

\tighten

{\renewcommand{\thefootnote}{\fnsymbol{footnote}}}
\setcounter{footnote}{0}


Charmed baryons play an important role in studies of strong and weak interactions, especially via investigations of their
semileptonic decays~\cite{slRMP,HQS,HQS2,xicBR_Theory1,xicBR_Theory2,xicBR_Theory3,xicBR_Theory4,xicBR_Theory5} and charge-parity violation (CPV)~\cite{cpvT,cpvH}. Such decay amplitudes are the product of a well-understood leptonic current for the lepton system and
a more complicated hadronic current for the quark transition. For semileptonic decays of
$SU(3)$ anti-triplets, $\Lambda^{+}_{c}$ and $\Xi_{c}^{+,0}$, thanks to the spin-zero light diquark constituents, a simpler and more powerful
theoretical calculation of form factors, hadronic structures, and nonperturbative aspects of strong interactions
can be performed in a relatively simple version of quantum chromodynamics (QCD)~\cite{slRMP}.

Thus far semileptonic decays of $\Lambda_{c}^{+}$ only have been comprehensively studied and are statistically limited by low production rates and/or high background levels of current experiments. Within uncertainties $CP$ symmetry and lepton flavor universality (LFU) are found to be conserved~\cite{lcbes1,lcbes2,lcAcp1,lcAcp2}. A violation of LFU would be a clear sign of new physics~\cite{LFU_theory1,LFU_theory2,LFU_theory3,LFU_theory4,LFU_theory5}. The tantalizing deviation from Standard Model predictions in $b$$\to$$c\ell\nu$ and $b$$\to$$s\ell\ell$ processes~\cite{RK_exp1,RK_exp2,RD_exp1,RD_exp2,Lb_LHCb,LU_LHCb,RD_sum} inspires tests of LFU in more semileptonic decays of heavy quarks.
For $\Xi^{0}_c$, the ARGUS collaboration first observed $18.1 \pm 5.9$ $\Xi_{c}^{0} \to \Xi \ell X$ events ($\ell=e$ or $\mu$)~\cite{arg_xic}. Later, the CLEO collaboration found $54 \pm 10$ $\Xi_{c}^{0} \to \Xi^{-} e^{+} \nu_{e}$ events~\cite{clo_xic}.
The ratio of the branching fractions, ${\cal B}(\Xi_{c}^{0} \to \Xi^{-} e^{+} \nu_{e})/{\cal B}(\Xi_{c}^{0} \to \Xi^{-} \pi^{+})$, was $0.96 \pm 0.43 \pm 0.18$ from ARGUS and $3.1 \pm 1.0 ^{+0.3}_{-0.5}$ from CLEO measurements, respectively.
With the absolute branching fraction ${\cal B}(\Xi_{c}^{0} \to \Xi^{-}\pi^{+}) = (1.80 \pm0.52)\%$ measured
by Belle recently~\cite{xic0_BR}, the averaged ${\cal B}(\Xi_{c}^{0} \to \Xi^{-} e^{+} \nu_{e})$ is $(2.34 \pm 1.59)\%$~\cite{PDG}. A variety of models have been developed to predict the decay branching fraction for ${\cal B}(\Xi_{c}^{0} \to \Xi^{-} e^{+} \nu_{e})$ resulting in a range from 1.35\% to $(7.26 \pm 2.54)$\%~\cite{xicBR_Theory1,xicBR_Theory2,xicBR_Theory3,xicBR_Theory4,xicBR_Theory5}. A precise measurement is crucial to test these models as well as to constrain the model parameters.

Though the Standard Model accommodates CPV which is one of the conditions needed to explain our	matter-dominated universe~\cite{cpv1}, the magnitude of this effect as predicted by the KM mechanism is not	sufficient~\cite{cpv2}. CPV has been established in many meson decays~\cite{Kcpv,Bcpv1,Bcpv2,Bcpv3,Bcpv4,Bcpv5,Bcpv6,Bcpv7,Dcpv}, but CPV has never been observed in any baryon system. Studies of $CP$-violating processes in the charm baryon sector are very scarce~\cite{lcAcp1,lcAcp2,lcAcp3,lcAcp4,lcAcp5}. Since there should be CPV sources other than those currently known, it is imperative to search for those also in the charm baryon sector, and several phenomenology studies about CPV in charmed baryon decays have been conducted~\cite{CPV_theory1,CPV_theory2,CPV_theory3,CPV_theory4}.

$CP$ violation in two body decays of charmed baryons can manifest itself as an asymmetry between the parity-violating decay parameter $\alpha$ for a process and its charge conjugate. For the $\Xi_{c}^{0}\to \Xi^{-} \pi^{+} \to \Lambda \pi^{-} \pi^{+}$ process, the decay parameter $\alpha_{\Xi^{-}\pi^{+}}$ (denoted as $\alpha^{+}$) enters the angular distribution expression,
\begin{equation}\label{fun1}
	\frac{dN}{d\cos\theta_{\Xi^{-}}} \propto 1 + \alpha_{\Xi^{-}\pi^{+}}\alpha_{\Xi^{-}}\cos\theta_{\Xi^{-}}.
\end{equation}
Here, $\theta_{\Xi^{-}}$ is the angle between the $\Lambda$ momentum vector and the opposite of the $\Xi_c^0$ momentum in the $\Xi^-$ rest frame~\cite{helicity}, $dN$ is the number of signal events in each $\cos\theta_{\Xi^{-}}$ bin, and $\alpha_{\Xi^{-}}$ is decay parameter of the $\Xi^{-}$~\cite{BES_alphaXi}.
The definition of $\alpha_{\bar{\Xi}^{+}\pi^{-}}$ (denoted as $\alpha^{-}$) is analogous for the charge-conjugated decay mode.
The only charge-averaged measurement of the decay parameters $\alpha_{\Xi \pi}$ is from CLEO with the result $-0.56 \pm 0.39^{+0.10}_{-0.09}$~\cite{clo_a_xic}, which falls in the range of [$-0.99, -0.38$] expected from theoretical predictions~\cite{a_theory1,a_theory2,a_theory3, a_theory4,a_theory5,a_theory6}. The $CP$ asymmetry parameter $\AW_{CP} = (\alpha^{+} + \alpha^{-})/(\alpha^{+} - \alpha^{-})$ can be calculated for $\Xi_{c}^{0} \to \Xi^{-} \pi^{+}$ and $\bar{\Xi}_{c}^{0} \to \bar{\Xi}^{+} \pi^{-}$.

In this Letter, we present measurements of the branching fractions of
$\Xi_{c}^{0} \to \Xi^{-} \ell^{+} \nu_{\ell}$~\cite{charge-conjugate} with significantly improved precision
using data samples of 89.5 and 711 fb$^{-1}$ collected at $\sqrt{s} = 10.52$ and $\sqrt{s} = 10.58$ GeV, respectively, by the
Belle detector~\cite{Belle} at the KEKB asymmetric-energy collider~\cite{KEKB}. LFU is tested using these measured results.
Charm baryons are produced in processes such as $\EE \to c\bar{c}  \to \Xi_{c}^{0} + anything$. $\Xi^{-}$ is reconstructed via the $\Lambda \pi^-$ mode, and $\Lambda$ decays into $p\pi^{-}$.
The decay parameters of $\alpha^+$ and $\alpha^-$ and the $CP$-asymmetry parameter $\AW_{CP}$ are first measured for $\Xi_{c}^{0} (\bar{\Xi}_{c}^{0}) \to \Xi \pi$.

To optimize the signal selection criteria and calculate the signal reconstruction efficiency, we use Monte Carlo (MC) simulated events.
The $\EE \to c\bar{c}$ process is simulated with {\sc pythia}~\cite{pythia}, while the signal events of $\Xi_c^0$ semileptonic decays are generated using form factors from Lattice QCD calculation~\cite{LQCD}, and $\Xi_{c}^{0} \to \Xi^{-} \pi^{+}$ decays are generated with {\sc EvtGen}~\cite{evtgen}.
The MC events are processed with a detector simulation based on {\sc geant3}~\cite{geant3}. Simulated $\Upsilon(4S)\to B \bar{B}$ events with $B=B^+$ or $B^0$, and $e^+e^- \to q\bar{q}$ events with $q=u,~d,~s,~c$
at $\sqrt{s}=10.52$ GeV and 10.58 GeV,
are used as background samples in which the signals are removed, which are called generic simulated samples.

For leptons and pions which are direct childs of $\Xi_{c}^{0}$, the impact parameters perpendicular to and along the $e^{+}$ beam direction with respect to the interaction point are required to be less than 0.5 cm and 4 cm, respectively, and transverse momentum is restricted to be higher than 0.1 GeV/$c$.
For charged tracks, information from different detector subsystems is
combined to form the likelihood $\mathcal{L}_{i}$ for species $(i)$, where $i= e,~\mu,~\pi$,~$K$, or $p$~\cite{pid}.
A track not from $\Lambda$ with a likelihood ratio $\mathcal{L}_{\pi}/(\mathcal{L}_K + \mathcal{L}_\pi)\textgreater 0.6$
is identified as a pion.
With this selection, the pion identification
efficiency is about 94\%, while 5\% of the kaons are misidentified as pions.
A track with a likelihood ratio $\mathcal{L}_e/(\mathcal{L}_e+\mathcal{L}_{{\rm non}-e}) \textgreater 0.9$
is identified as an electron~\cite{lke}. The $\gamma$
conversions are removed by examining all combinations of an $e^{\pm}$ track with other oppositely-charged tracks in the event that are identified as $e^{\mp}$, and requiring $\EE$ invariant mass larger than $0.2$ GeV/$c^{2}$.
Tracks with $\mathcal{L}_{\mu}/(\mathcal{L}_{\mu}+\mathcal{L}_{K}
+\mathcal{L}_{\pi})\textgreater0.9$ are considered as muon candidates~\cite{lku}.
Furthermore, the muon tracks are required to hit at least five layers of the $K^{0}_{L}$ and muon subdetector, and not to be identified as kaons
with $\mathcal{L}_{K}/(\mathcal{L}_K + \mathcal{L}_\pi)< 0.4$ to suppress backgrounds due to misidentification. With the above selections,
the efficiencies of electron and muon identification are 96\% and 75\%, respectively, with pion fake rates less than 2\%.

Candidate $\Lambda$ baryons are reconstructed in the decay $\Lambda \to p \pi^-$
and selected if $|M_{p \pi^-}-m_{\Lambda}|<3$ MeV/$c^2$ ($\sim2.5\sigma$), where $\sigma$ denotes the mass resolution. Here and throughout the text,
$M_i$ represents a measured invariant mass and $m_i$ denotes the
nominal mass of the particle $i$~\cite{PDG}.
The proton track from $\Lambda$ decay is required to satisfy $\mathcal{L}_{p}/(\mathcal{L}_{\pi} + \mathcal{L}_p)\textgreater 0.2$ and $\mathcal{L}_{p}/(\mathcal{L}_{K} + \mathcal{L}_p)\textgreater 0.2$ with an efficiency of 95\%.
We define the $\Xi^-$
signal region as $|M_{\Lambda\pi^-}-m_{\Xi^-}|<6.5$ MeV/$c^{2}$ ($\sim3 \sigma$),
and $\Xi^-$ mass sidebands as $1.294$ GeV/$c^2$ $< M_{\Lambda\pi^-} <$ 1.307 GeV/$c^2$
and $1.337$ GeV/$c^2$ $< M_{\Lambda\pi^-} < 1.350$ GeV/$c^2$. To suppress combinational background, we require
the flight directions of $\Lambda$ and $\Xi^{-}$ candidates, which are reconstructed from their fitted production and decay vertices, to be within five degrees of their momentum directions. We also require the scaled momentum $p^{*}_{\Xi^{-} X}/p^{*}_{\rm max} \textgreater 0.45$
($X = e^{+},~\mu^{+}$ or $\pip$), where $p^{*}_{\Xi^{-} X}$ is the momentum of the $\Xi^{-} X$ system in the center-of-mass frame
and $p^{*}_{\rm max} \equiv \sqrt{E_{\rm beam}^{2} - m^{2}_{\Xi_{c}^{0}}}$ ($E_{\rm beam}$ is the beam energy). This requirement removes all $\Xi_{c}^{0} \to \Xi^{-} \pi^{+}$ decays with $\Xi_{c}^{0}$ produced in $B$ decays from the $\sqrt{s}=10.58$ GeV sample. For $\Xi_{c}^{0} \to \Xi^{-} \ell^{+} \nu_{\ell}$, the cosine of the opening angle
between $\Xi^{-}$ and $\ell^{+}$ is further required to be larger than 0.25.


\begin{figure*}[htbp]
	\begin{center}
		\includegraphics[width=3.7cm]{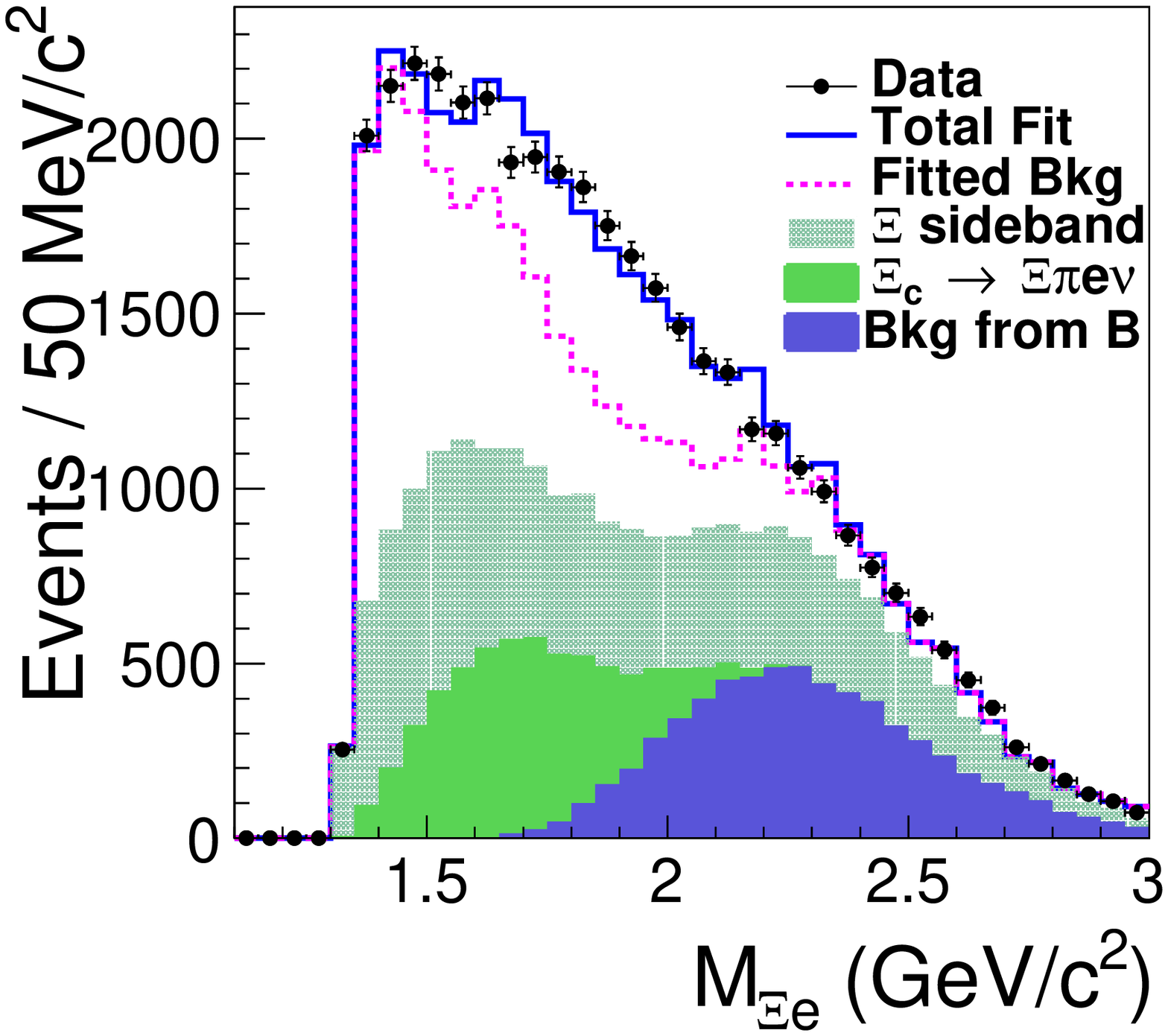}
		\includegraphics[width=3.7cm]{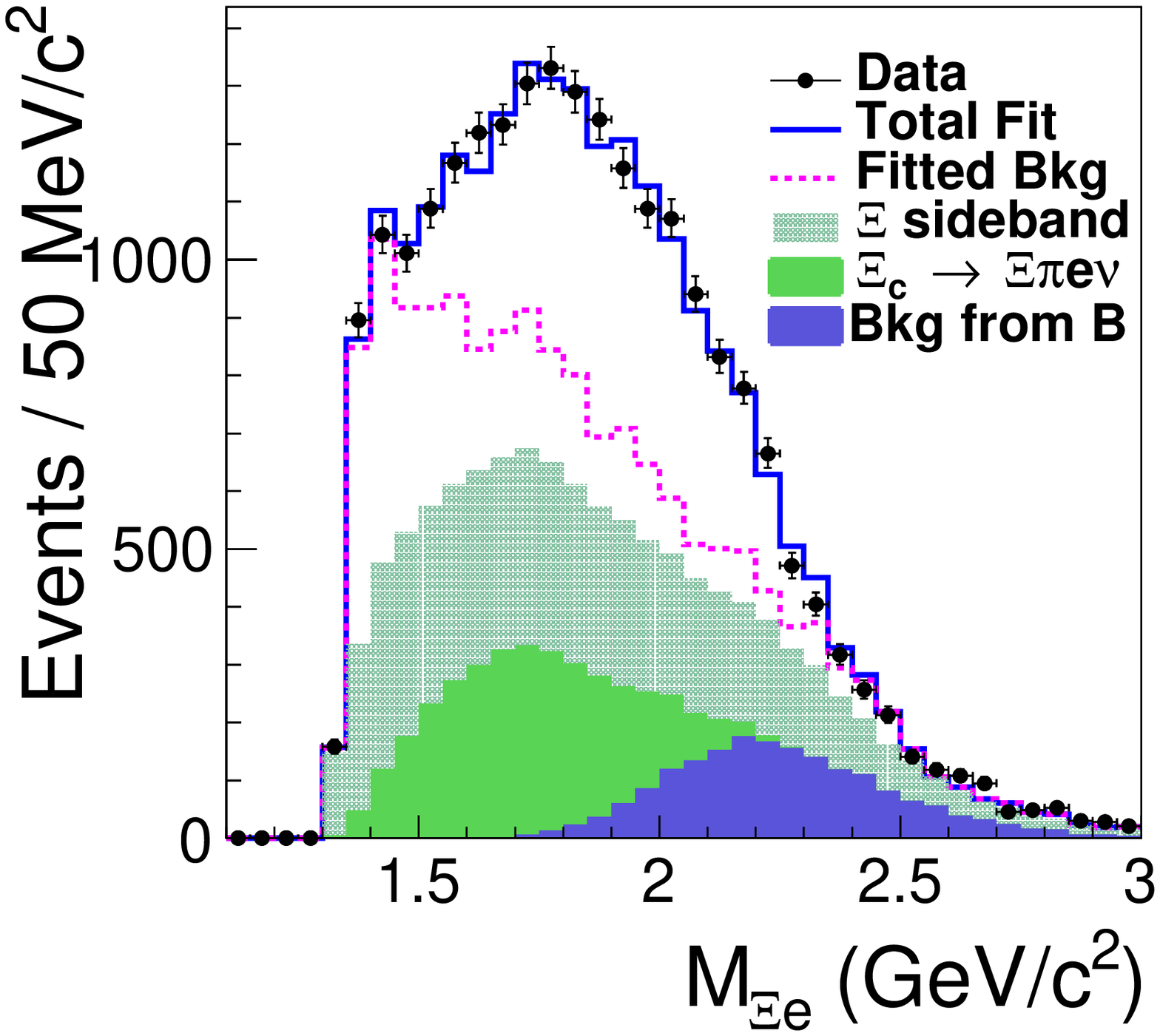}
		\includegraphics[width=3.7cm]{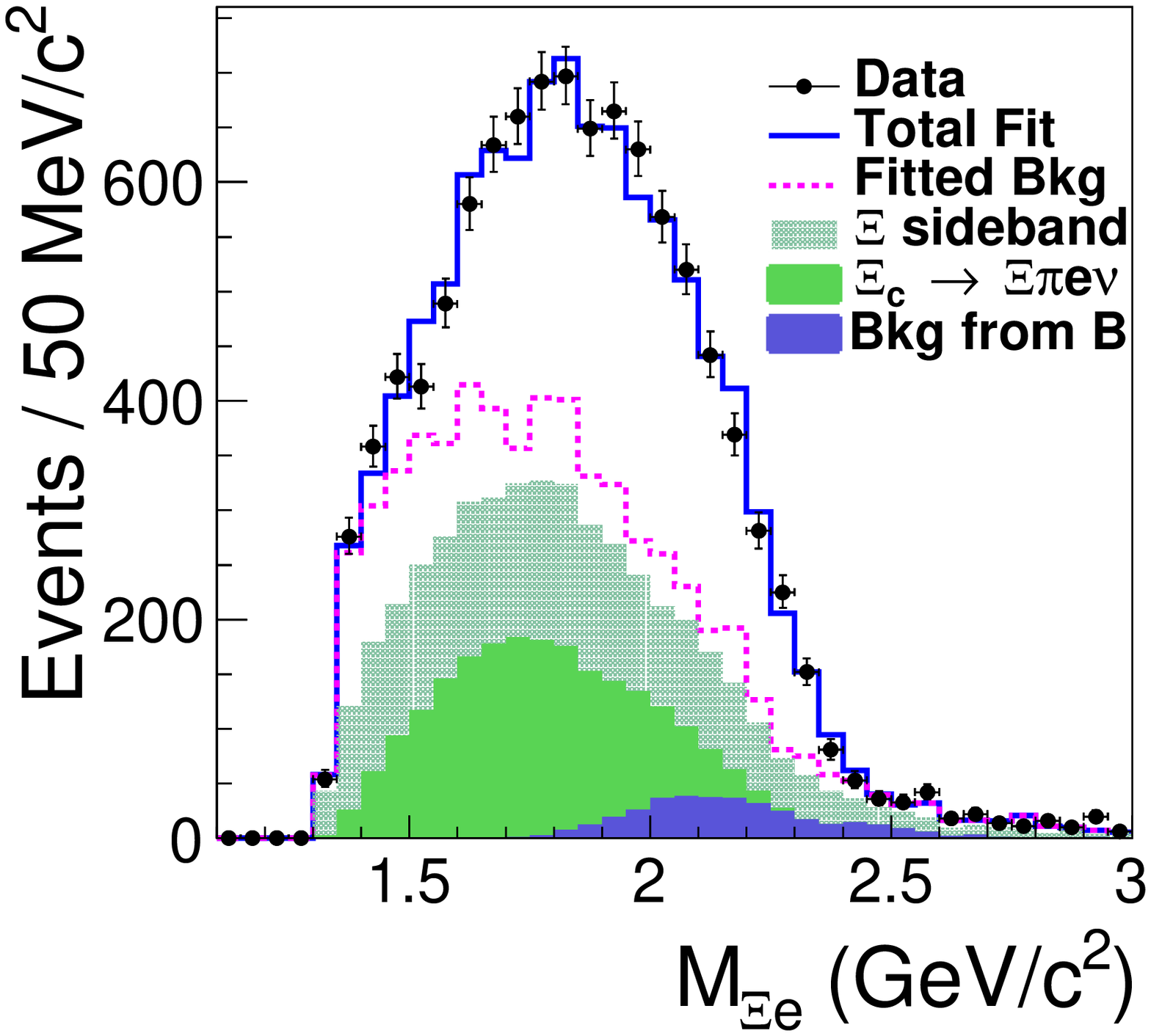}
		\includegraphics[width=3.7cm]{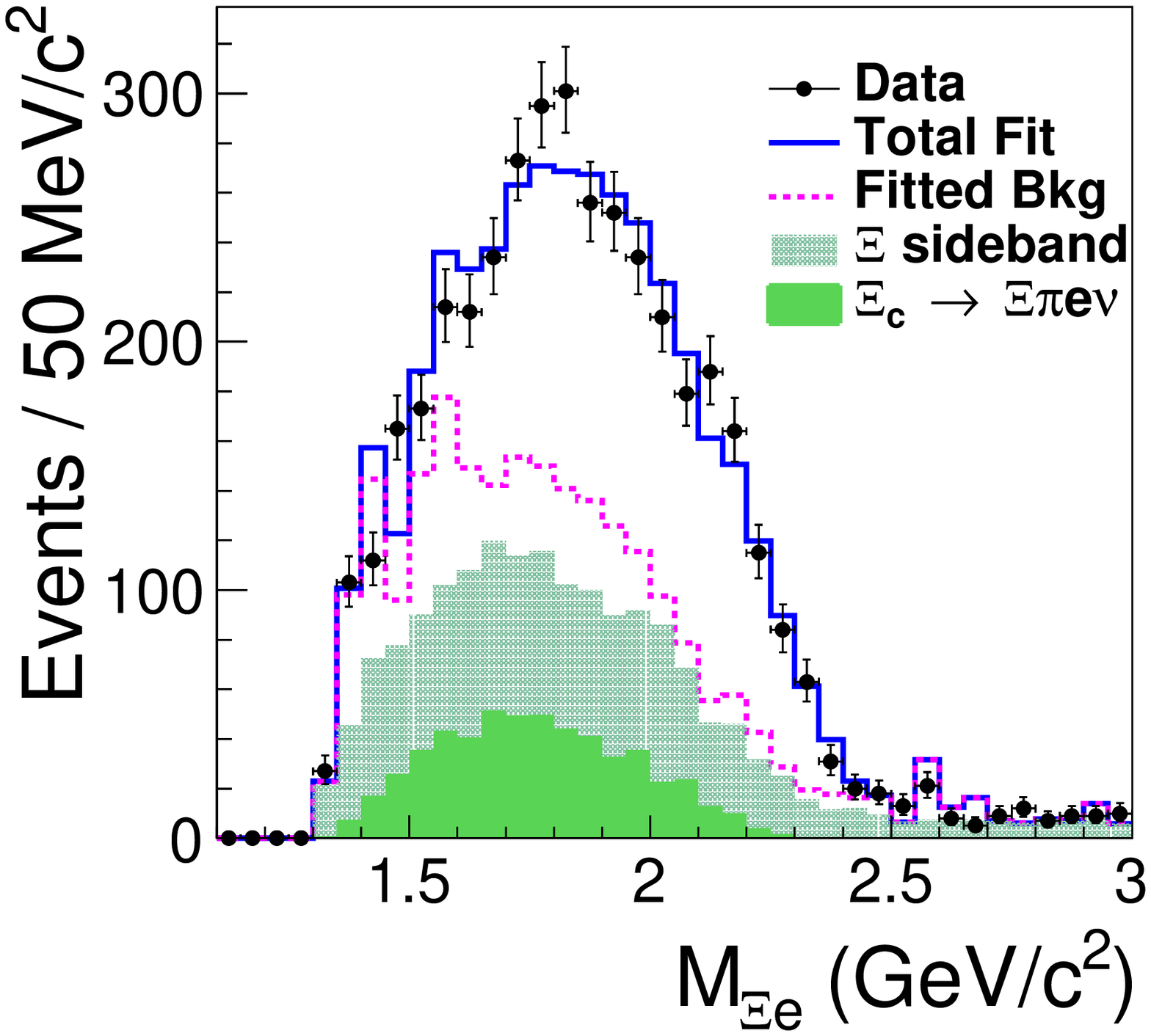}
		\put(-450,45){\bf (a):}
		
		\includegraphics[width=3.7cm]{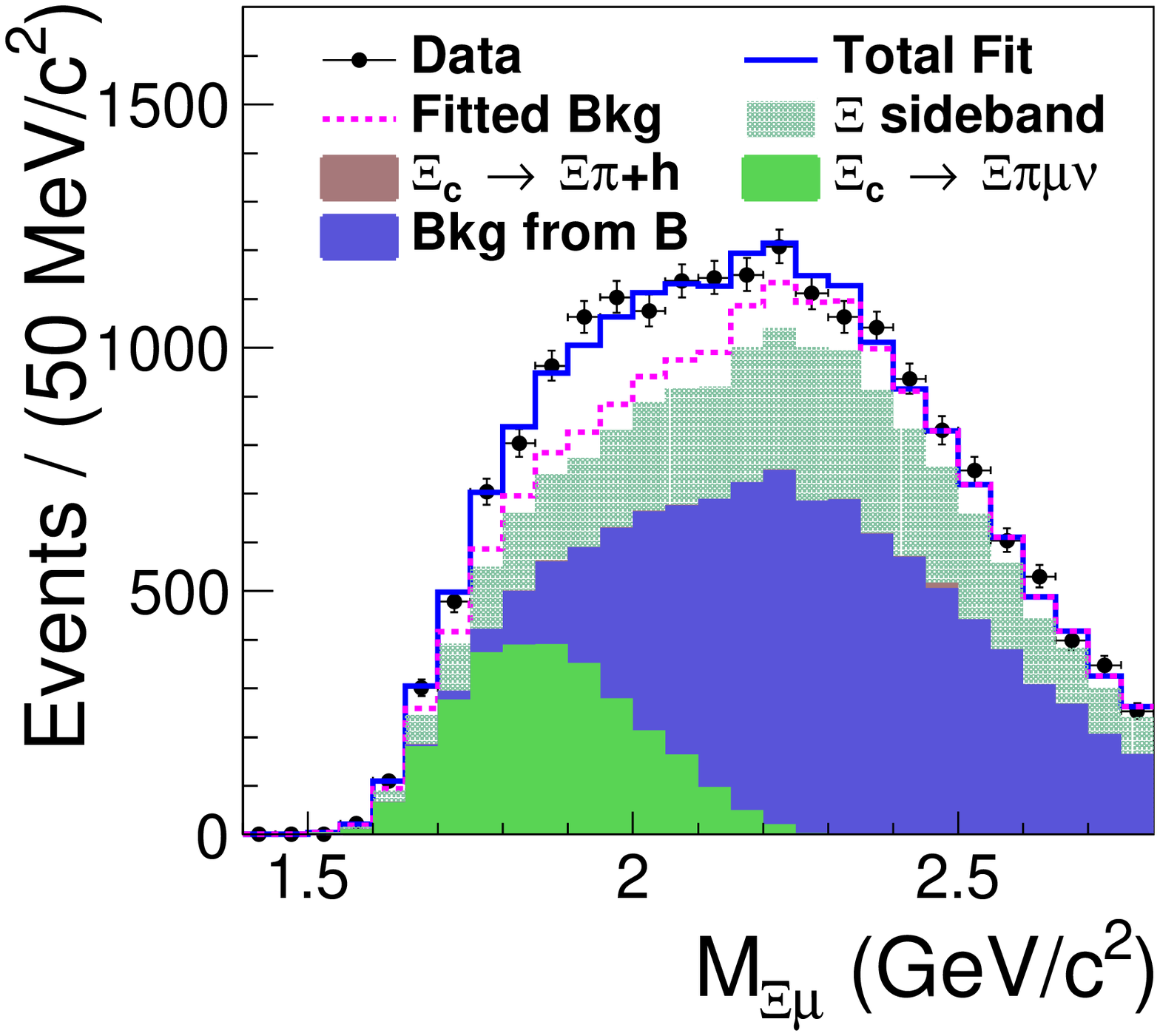}
		\includegraphics[width=3.7cm]{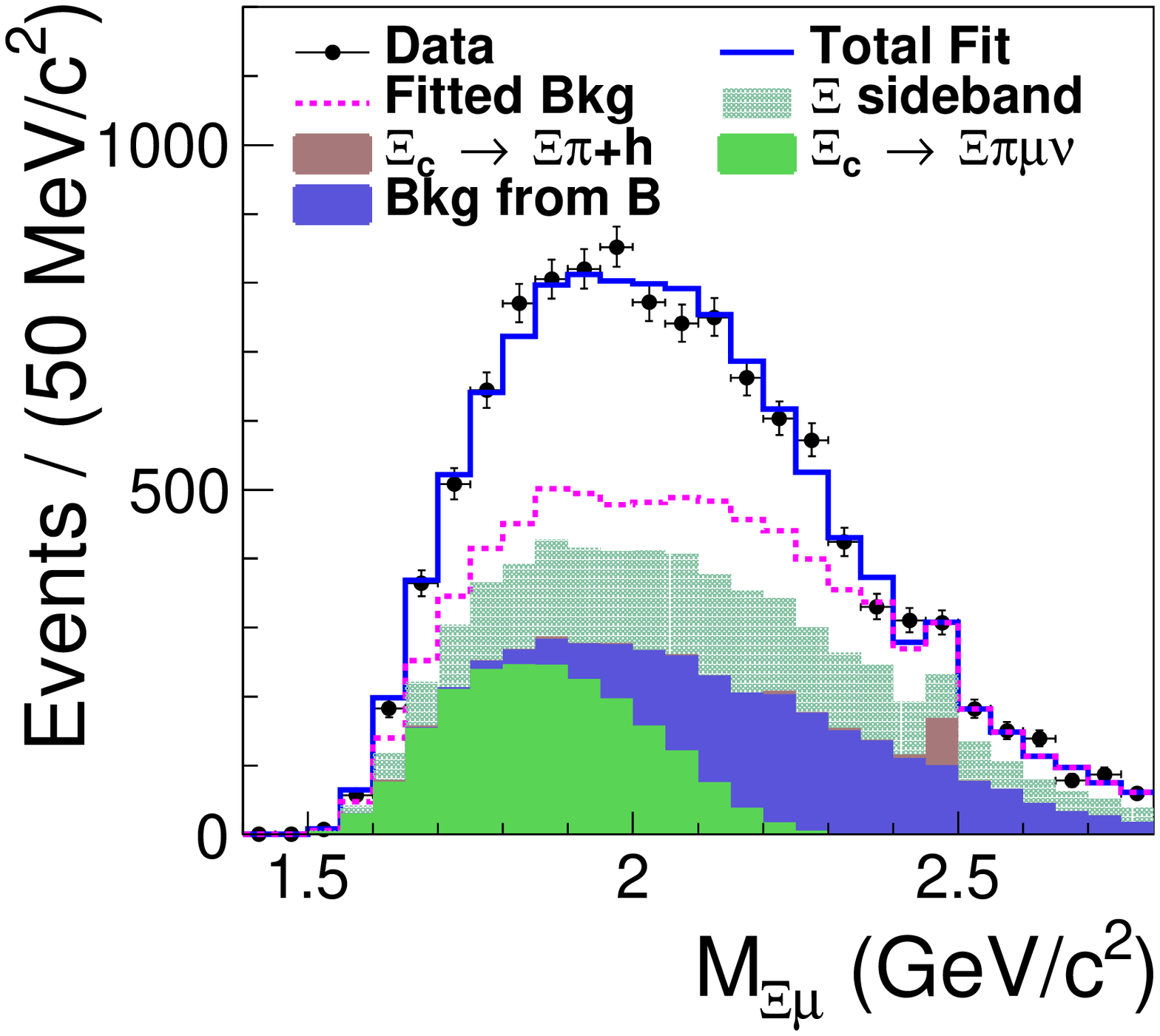}
		\includegraphics[width=3.7cm]{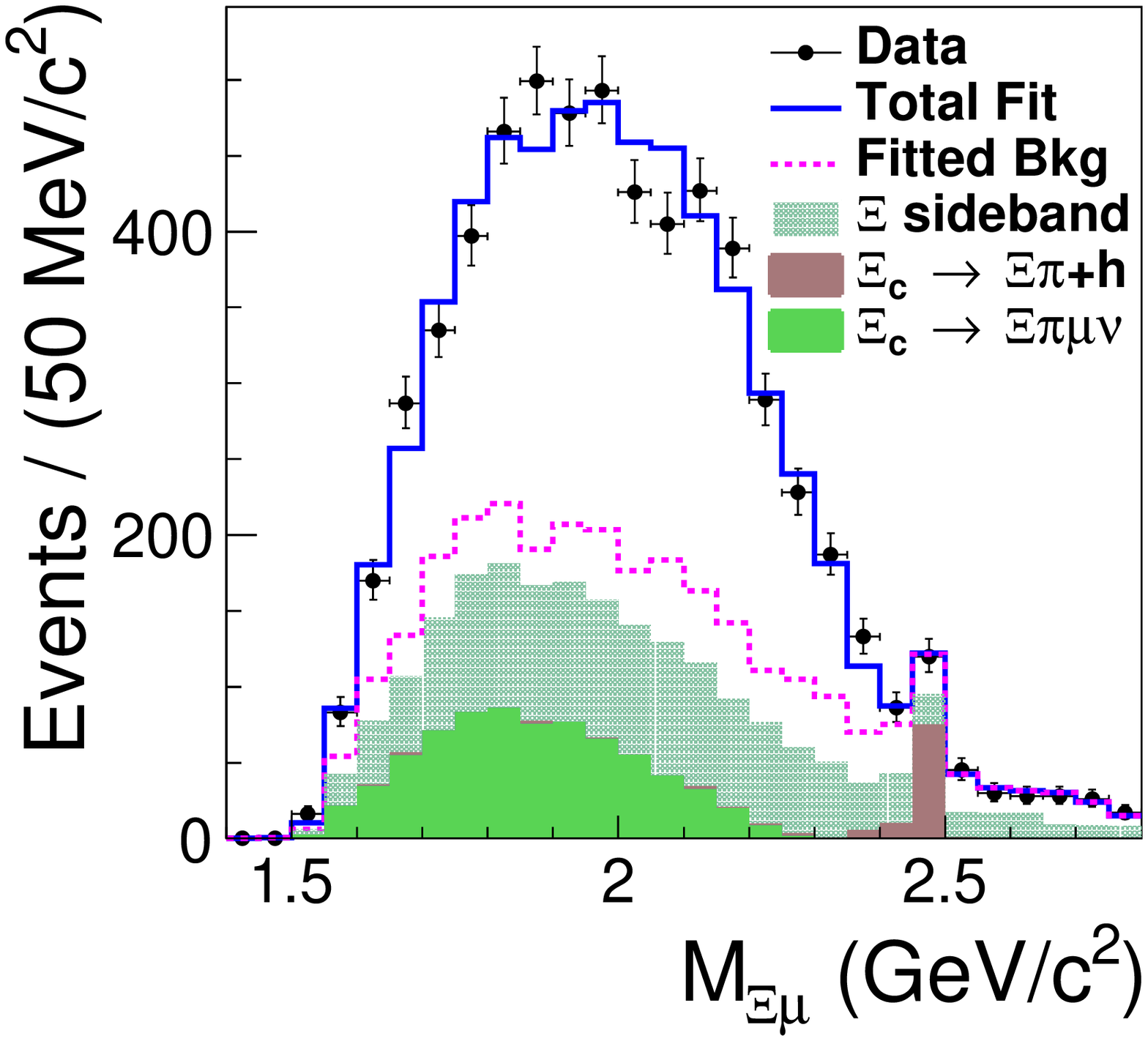}
		\includegraphics[width=3.7cm]{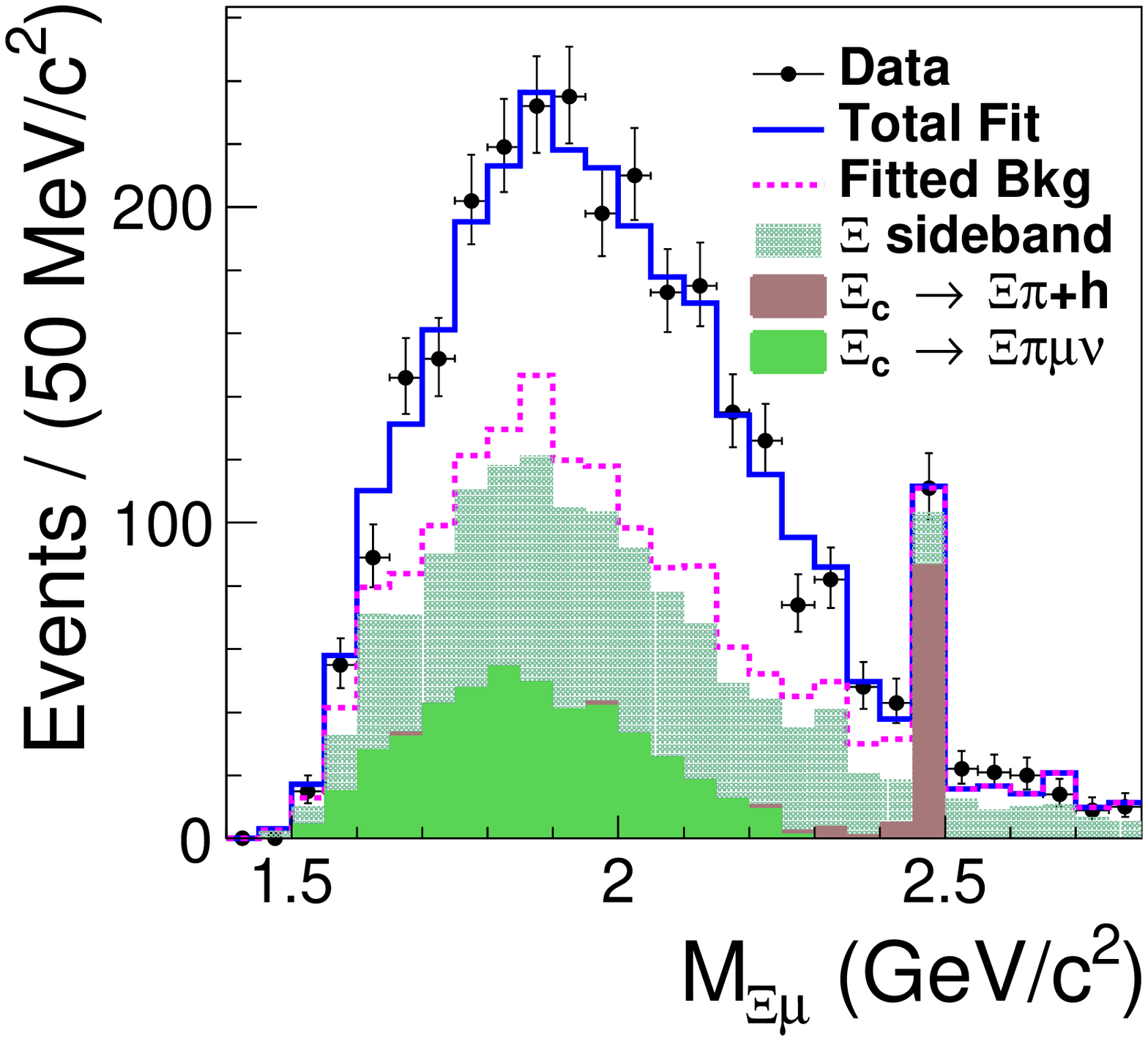}
		\put(-450,45){\bf (b):}
		
		\includegraphics[width=3.7cm]{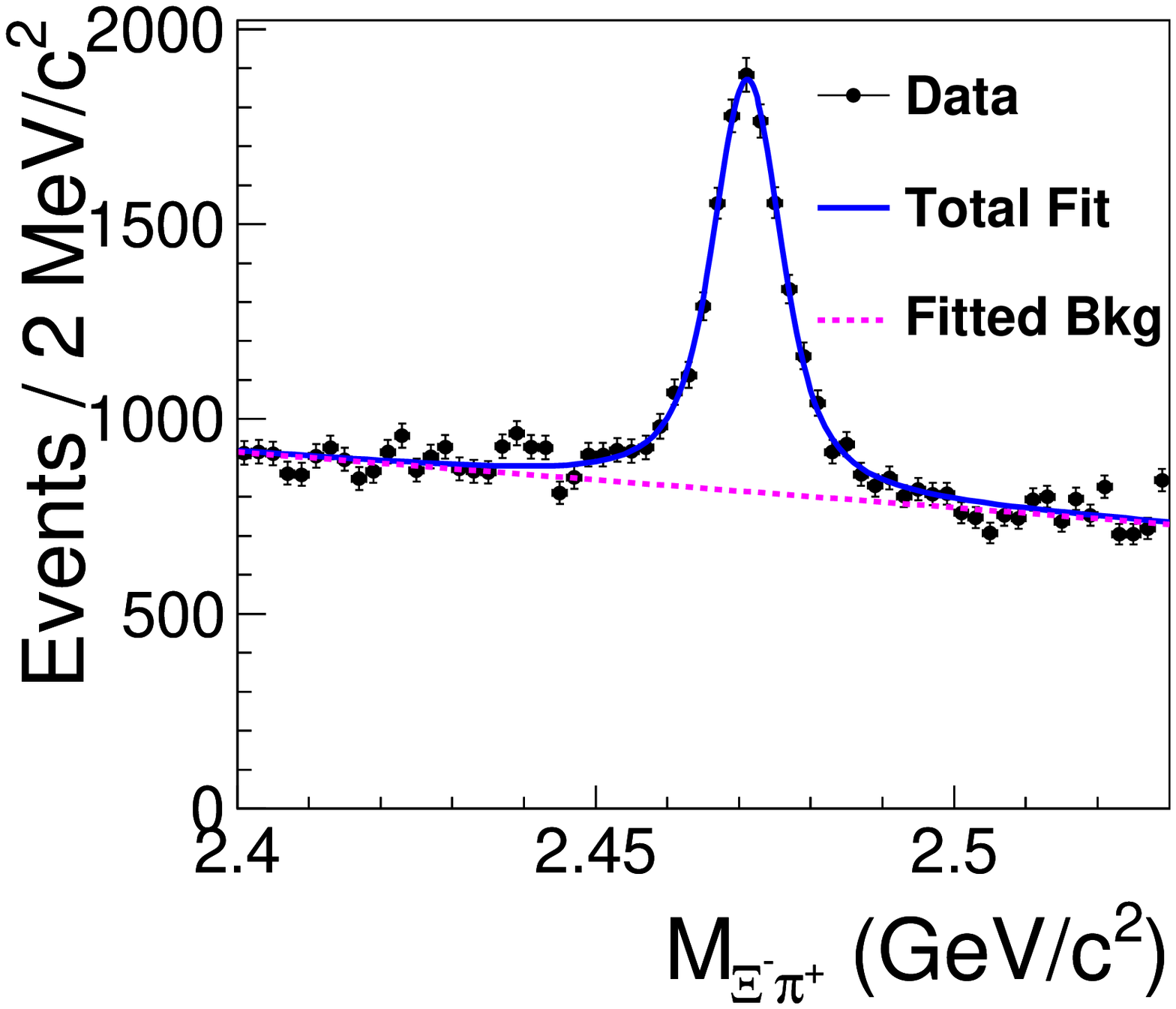}
		\includegraphics[width=3.7cm]{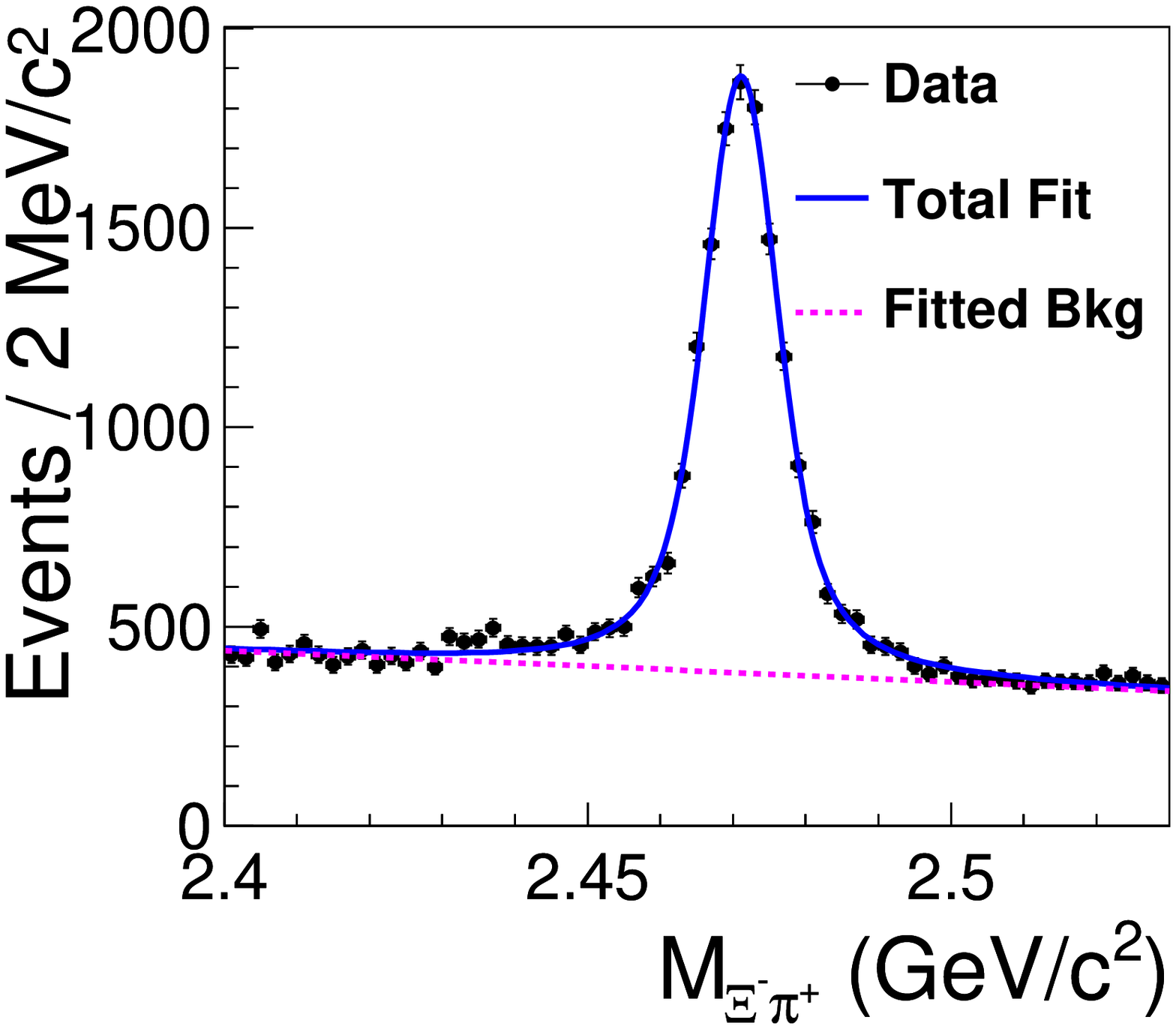}
		\includegraphics[width=3.7cm]{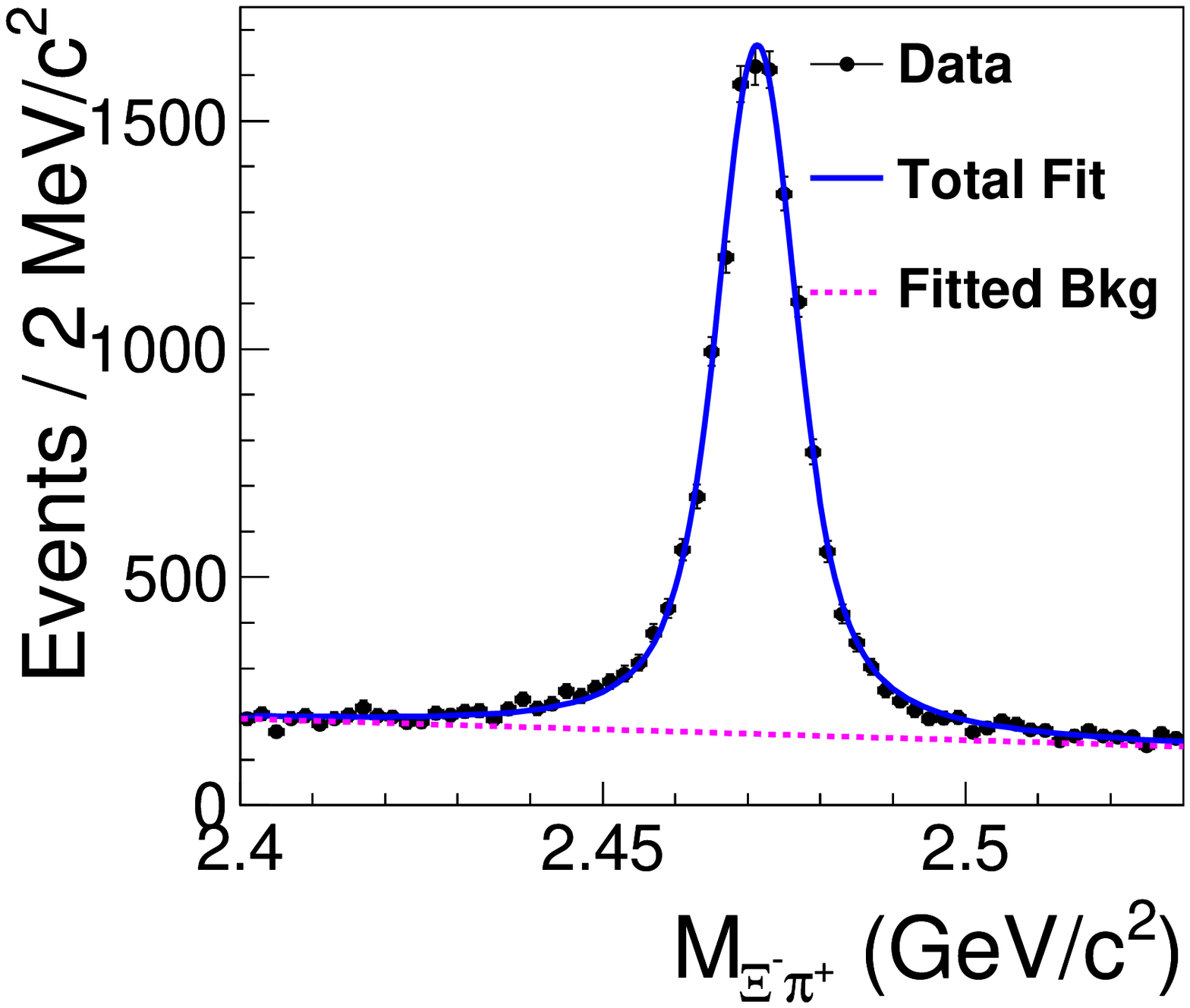}
		\includegraphics[width=3.7cm]{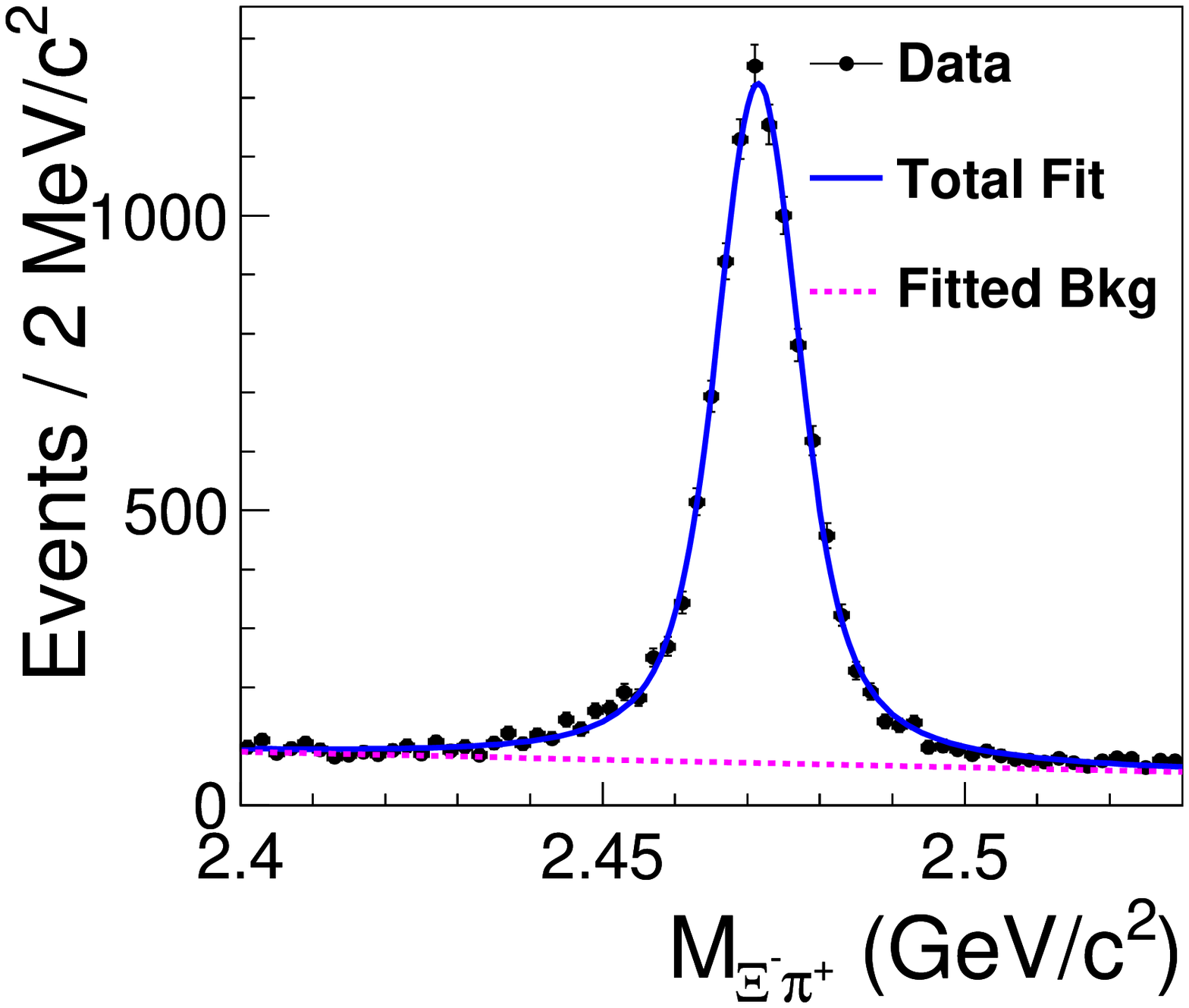}
		\put(-450,45){\bf (c):}
		\put(-490,-10) { $p^{*}_{\Xi^{-} X}/p^{*}_{\rm max}$ region: } \put(-400,-10){ (0.45, 0.55) } \put(-290,-10){ (0.55, 0.65) } \put(-180,-10){ (0.65, 0.75) } \put(-65,-10){(0.75,1)}
		\caption{ The fits to the $M_{\Xi^{-} e^{+}}$, $M_{\Xi^{-} \mu^{+}}$, and $M_{\Xi^{-} \pi^{+}}$ distributions of the selected
(a) $\Xi_{c}^{0} \to \Xi^{-} e^{+} \nu_{e}$, (b) $\Xi_{c}^{0} \to \Xi^{-} \mu^{+} \nu_{\mu}$, and (c) $\Xi_{c}^{0} \to \Xi^{-} \pi^{+}$
candidates in each $p^{*}_{\Xi^{-} X}/p^{*}_{\rm max}$ bin listed at the bottom. The points with error bars represent the data
from $\sqrt{s}=10.52$ GeV and 10.58 GeV, the solid blue lines are the best fits, and the violet dashed lines are the fitted total backgrounds. The other components of the fits are indicated in the legends.}\label{fit_br}
	\end{center}
\end{figure*}

After the above selections, the obtained $\Xi^{-} e^{+}$, $\Xi^{-} \mu^{+}$, and $\Xi^{-} \pi^{+}$
mass spectra from data in $p^{*}_{\Xi^{-} X}/p^{*}_{\rm max}$ regions of (0.45, 0.55), (0.55, 0.65), (0.65, 0.75), and
(0.75, 1) are shown in Fig.~\ref{fit_br}.
The $\Xi_{c}^{0}$ signals are extracted from maximum-likelihood fits to
these invariant mass spectra. For $\Xi_{c}^{0}$ semileptonic decays, the signal shapes are taken directly from MC simulation. The background shapes from wrongly constructed $\Xi$ candidates can be described by the $M_{\Xi^{-} \ell^{+}}$ distributions of $\Xi^{-}$ mass sidebands. The backgrounds from $\Xi_{c} \to \Xi \pi \ell^{+} \nu_{\ell}$ are taken from MC simulations of those processes. The backgrounds from $e^+ e^- \to q\bar{q}$ due to mis-selected $\ell^{+}$ can be represented by the $M_{\Xi^{-} \ell^{+}}$ distributions of $\Xi^{-}\ell^{-}$ events with their normalized $\Xi^{-}$ mass sidebands subtracted. The other backgrounds are from $e^+ e^- \to B\bar{B}$ with $\Xi^{-}$ from one $B$ and $\ell^{+}$ from another $\bar{B}$, whose shapes are taken from generic simulated samples. Background from $\Omega_{c}^{0} \to \Xi^{-} \ell^{+} \nu_{\ell}$ decays is assumed to be negligible since it is a $c$ $\to$ $d$ process and should be suppressed strongly. In fitting the $\Xi^{-} \mu^{+}$ mass spectrum, an additional background of simulated $\Xi_{c}^{0,+} \to \Xi^{-} \pi^{+}$$+$$hadrons$ events from generic simulated samples is added. In the fit above, the shapes of all fit components are fixed while their yields are floated.
In fitting the $\Xi^{-} \pi^{+}$ mass spectrum, the $\Xi_{c}^{0}$ signal shape is parameterized with a double-Gaussian function with same mean value and all other parameters floated,
while the background shape is represented with a 1st-order polynomial.
Figure~\ref{fit_br} shows the fitted results in each $p^{*}_{\Xi^{-} X}/p^{*}_{\rm max}$ bin labelled at the bottom for (a) $\Xi_{c}^{0} \to \Xi^{-} e^{+} \nu_{e}$, (b) $\Xi_{c}^{0} \to \Xi^{-} \mu^{+} \nu_{\mu}$, and (c) $\Xi_{c}^{0} \to \Xi^{-} \pi^{+}$.
The fitted result in each $p^{*}_{\Xi^{-} X}/p^{*}_{\rm max}$ bin together with the corresponding detection efficiency
are listed in Table~\ref{tab:br}. The background sources and fit methods
are validated with generic simulated samples.

\begin{table*}[htbp]
	\caption{\label{tab:br} List of the fitted signal yields and the corresponding
detection efficiencies in each $p^{*}_{\Xi^{-} X}/p^{*}_{\rm max}$ bin ($N^{\Xi^{-}X}_{i}/\varepsilon^{\Xi^{-}X}_{i}$) of data at $\sqrt{s}=10.52$ GeV and 10.58 GeV.
The last column gives the ratios of branching fractions $\frac{{\cal B}(\Xi_{c}^{0} \to \Xi^{-}\ell^{+} \nu_{\ell})}{{\cal B}(\Xi_{c}^{0} \to \Xi^{-} \pi^{+})}$
in the full $p^{*}_{\Xi^{-} X}/p^{*}_{\rm max}$ range. Quoted uncertainties are statistical only.}
	\begin{center}
		\renewcommand\arraystretch{2.1}
		\resizebox{\textwidth}{!}{
			\begin{tabular}{cccccc}
				\toprule[2pt]
				$p^{*}_{f}/p^{*}_{\rm max}$  &       $(0.45,0.55)$           &           $(0.55,0.65)$           &           $(0.65,0.75)$           &           $ > 0.75$          & $\frac{{\cal B}(\Xi_{c}^{0} \to \Xi^{-}\ell^{+} \nu_{\ell})}{{\cal B}(\Xi_{c}^{0} \to \Xi^{-} \pi^{+})}$ \\
				\hline
				$\Xi_{c}^{0} \to \Xi^{-} e^{+} \nu_{e}$ & $(5.13 \pm 0.26 )\times10^{3}/16.5\%$  & $(6.08 \pm 0.28) \times10^{3}/19.8\%$  & $(4.08 \pm 0.21 ) \times10^{3}/21.4\%$ & $(1.72 \pm 0.10) \times10^{3}/21.3\%$ &       $0.730 \pm 0.021 $  \\
				$\Xi_{c}^{0} \to\Xi^{-} \mu^{+} \nu_{\mu}$&  $(1.68 \pm 0.15)\times10^{3}/6.59\%$  & $(3.35 \pm 0.25) \times10^{3}/10.36\%$  & $(3.15 \pm 0.25) \times10^{3}/13.6\%$  & $(1.13\pm 0.14) \times10^{3}/15.5\%$ &    $0.708\pm 0.033  $   \\
				$\Xi_{c}^{0} \to\Xi^{-} \pi^{+}$       &     $(8.12 \pm 0.29)\times 10^{3}/24.2\% $ & $(1.20 \pm 0.02) \times 10^{4}/25.6\%$ & $(1.31 \pm 0.02)\times 10^{4}/26.9\%$  & $(1.04 \pm	0.02)\times 10^{4}/27.13\%$ &                    $...$         \\
				\toprule[2pt]
		        \end{tabular}
	        	    }
	\end{center}
\end{table*}

The $\Xi_{c}^{0}$ semileptonic decay branching fractions are calculated using
$$
{\cal B}(\Xi_{c}^{0} \to \Xi^{-}\ell^{+} \nu_{\ell}) \equiv \frac{\varepsilon_{\rm pop}^{\Xi^{-}\pi^{+}} \Sigma_{i}\frac{N^{\Xi^{-}\ell^{+}}_{i}}{\varepsilon^{\Xi^{-}\ell^{+}}_{i}}}{\varepsilon_{\rm pop}^{\Xi^{-}\ell^{+}}\Sigma_{i}\frac{N^{\Xi^{-}\pi^{+}}_{i}}{\varepsilon^{\Xi^{-}\pi^{+}}_{i}}} \times {\cal B}(\Xi_{c}^{0} \to \Xi^{-} \pi^{+}),
$$
where $N^{\Xi^{-}X}_{i}$ and $\varepsilon^{\Xi^{-}X}_{i}$ are the fitted signal yield and detection efficiency, respectively, in each $p^{*}_{\Xi^{-}X}/p^{*}_{\rm max}$ bin; $\varepsilon_{\rm pop}^{\Xi^{-}X}$ is the efficiency of the
$p^{*}_{\Xi^{-}X}/p^{*}_{\rm max} \textgreater 0.45$ requirement for each channel and is 0.783, 0.574, and 0.588
for $\Xi_{c}^{0} \to \Xi^{-} \pi^{+}$, $\Xi^{-} e^{+} \nu_{e}$, and $\Xi^{-} \mu^{+} \nu_{\mu}$, respectively.

Using the results listed in Table~\ref{tab:br}, we obtain ${\cal B}(\Xi_{c}^{0} \to \Xi^{-} e^{+} \nu_{e}) = (1.31 \pm 0.04 \pm 0.38)\%$,
${\cal B}(\Xi_{c}^{0} \to \Xi^{-} \mu^{+} \nu_{\mu}) = (1.27 \pm 0.06 \pm 0.37)\%$, and ${\cal B}(\Xi_{c}^{0} \to \Xi^{-} e^{+} \nu_{e})/{\cal B}(\Xi_{c}^{0} \to \Xi^{-} \mu^{+} \nu_{\mu}) = 1.03 \pm 0.05$. Here, the first and second uncertainties are statistical and from ${\cal B}(\Xi_{c}^{0} \to \Xi^{-}\pi^{+})$~\cite{xic0_BR}, respectively.

\begin{figure}[htbp]
	\begin{center}
		\includegraphics[width=4.3cm]{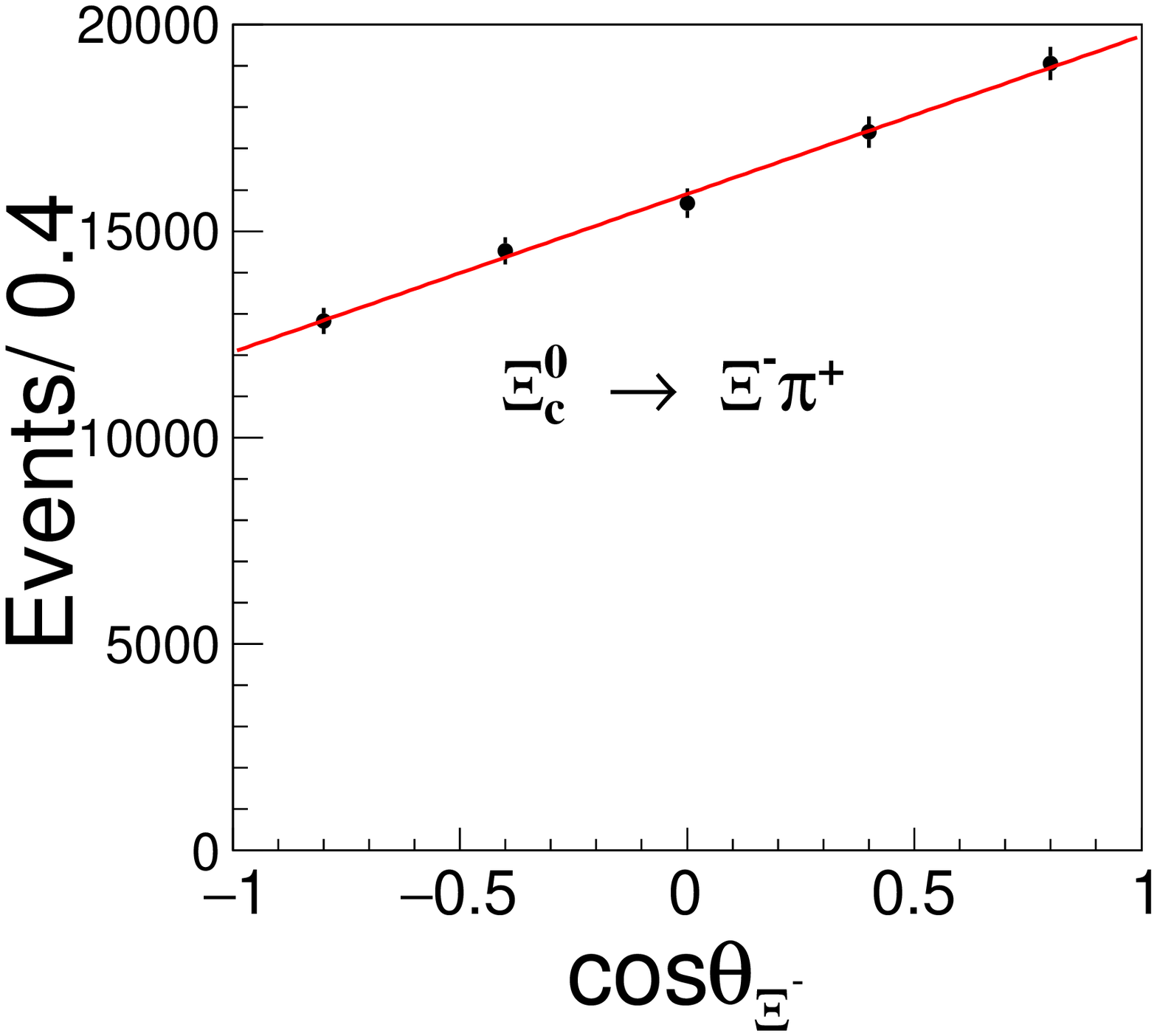}
		\includegraphics[width=4.3cm]{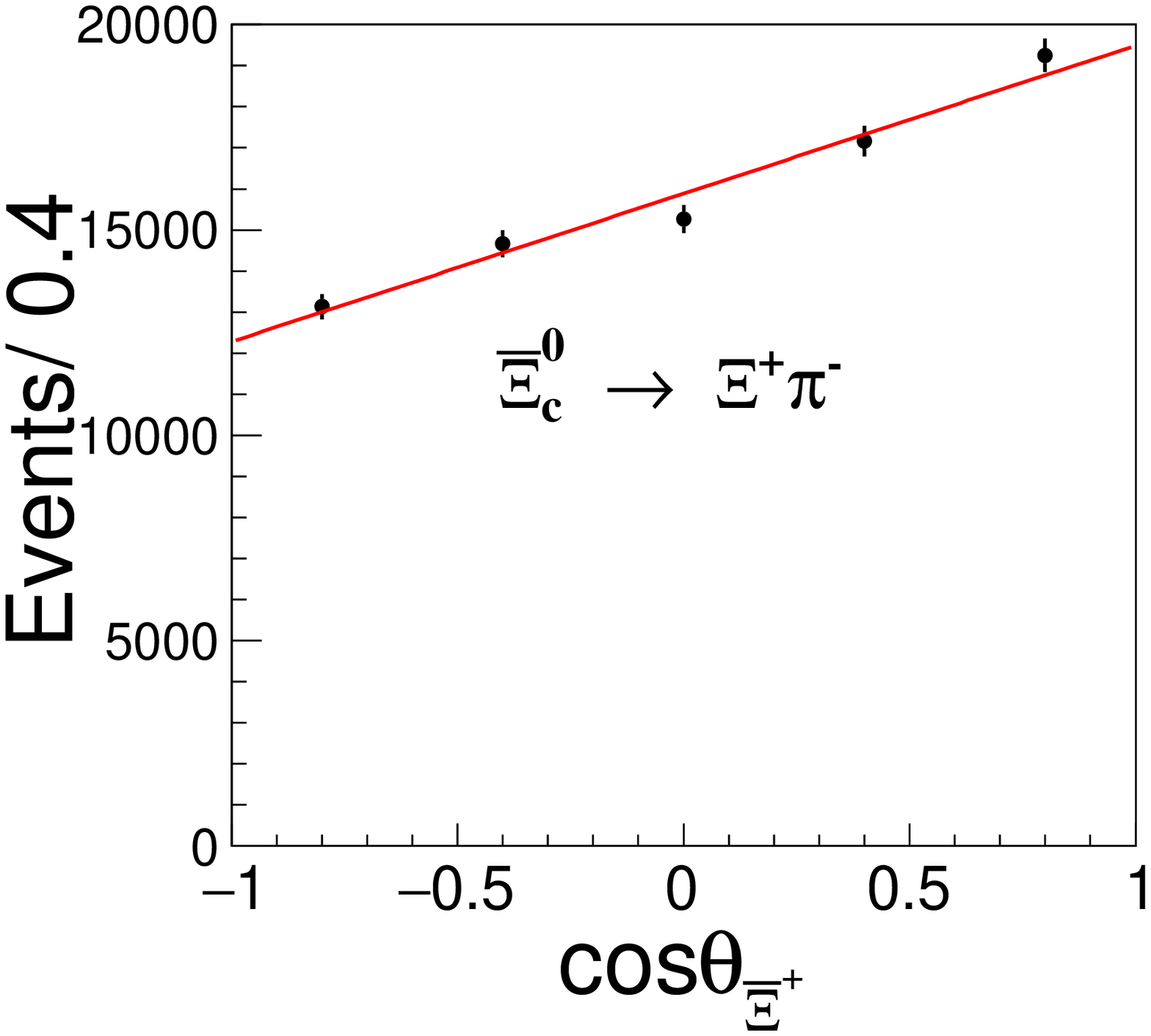}
		\put(-217,45){\bf (a): $\chi^{2}/ndf = 0.3$} 		\put(-93,45){\bf (b): $\chi^{2}/ndf = 1.8$}
		\caption{The maximum likelihood fits to the efficiency-corrected $\cos\theta_{\Xi}$ distributions of data to extract (a) $\alpha_{\Xi^{-}\pi^{+}}$ and (b) $\alpha_{\bar{\Xi}^{+}\pi^{-}}$
for $\Xi_{c}^{0}\to \Xi^{-} \pi^{+}$ and $\bar{\Xi}_{c}^{0}\to \bar{\Xi}^{+} \pi^{-}$ decays. The points with error bars represent data from the combined samples at $\sqrt{s}=10.52$ GeV and 10.58 GeV, and the red solid lines are the best fits. }\label{fit_alpha_final}
	\end{center}
\end{figure}

In the following, $\Xi_{c}^{0}\to \Xi^{-} \pi^{+}$ and $\bar{\Xi}_{c}^{0}\to \bar{\Xi}^{+} \pi^{-}$ decays are treated separately to
extract decay parameters of $\alpha^+$ and $\alpha^-$, and $\AW_{CP}$ for $\Xi_{c}^{0} (\bar{\Xi}_{c}^{0}) \to \Xi^{-} \pi^{+}(\bar{\Xi}^{+} \pi^{-})$. To obtain the $\theta_{\Xi}$ distribution, we divided the 2D plane of
$p^{*}_{\Xi \pi}/p^{*}_{\rm max}$ versus $\cos\theta_{\Xi}$ into $4 \times 5$ bins with the bin edges for $p^{*}_{\Xi \pi}/p^{*}_{\rm max}$ and $\cos\theta_{\Xi}$
set as $(0.45,~0.55,~0.65,~0.75,~1.0)$ and $(-1.0,~-0.6,~-0.2,~0.2,~0.6,~1.0)$, respectively. The detection efficiency in each 2D bin is calculated individually. The number of
$\Xi_{c}^{0} (\bar{\Xi}_{c}^{0})$ signal events in each 2D bin is obtained by fitting the corresponding
$M_{\Xi\pi}$ distribution with the method used in the branching fraction measurements.
The number of signal events in each $\cos\theta_{\Xi}$ bin is the sum of the efficiency-corrected signal yields
in corresponding $p^{*}_{\Xi \pi}/p^{*}_{\rm max}$ bins. The fitting method was checked using special simulated samples with a range of values of $\AW_{CP}$. 
The final efficiency-corrected $\cos\theta_{\Xi}$ distributions for (a) $\Xi_{c}^{0}\to \Xi^{-} \pi^{+}$ and (b) $\bar{\Xi}_{c}^{0}\to \bar{\Xi}^{+} \pi^{-}$ decays
are shown in Fig.~\ref{fit_alpha_final}. Using Eq.~(\ref{fun1}) with $\alpha_{\Xi^{-}} = -0.376 \pm 0.008$ and $\alpha_{\bar{\Xi}^{+}}= 0.371 \pm 0.007$~\cite{BES_alphaXi}, the fits yield $\alpha^{+} = -0.64 \pm 0.05 $ and $\alpha^{-} = 0.61 \pm 0.05$, resulting in $\AW_{CP}=0.024 \pm 0.052$.
Here, the uncertainties are statistical only. 

There are several sources of systematic uncertainties contributing to the
branching fraction measurements. Using the $D^{*+} \to D^{0}
 \pi^{+}$, $D^{0} \to K^{-} \pi^{+}$, $\Lambda \to p \pi$, and $J/\psi \to \ell \ell$ control samples, the particle identification
uncertainties ($\sigma_{\rm PID}$) are $0.51\%-0.55\%$ per pion,
$0.55\%-0.93\%$ per electron, and $0.44\%-0.84\%$ per muon, depending on the
$p^{*}_{\Xi^{-} X}/p^{*}_{\rm max}$ region.
The systematic uncertainties associated with tracking efficiency and $\Xi^{-}$ selection cancel in the branching fraction ratio measurements.
We estimate the systematic uncertainties associated with the fitting procedures ($\sigma_{\rm fit}$) for $\Xi_{c}^{0} \to \Xi^{-} \ell^{+} \nu_{\ell}$ and
$\Xi_{c}^{0}\to \Xi^{-} \pi^{+}$ separately.
For $\Xi_{c}^{0} \to \Xi^{-} \ell^{+} \nu_{\ell}$ decays, we change the bin width of the $M_{\Xi^{-} \ell^{+}}$ spectra by $\pm$5 MeV/$c^{2}$,
change the $\Xi^{-}$ mass sidebands from two times that of the signal region to three times that of the signal region, add the background component from $\Xi_{c} \to \Xi \pi^{+} \pi^{-} \ell^{+} \nu_{\ell}$ with its shape taken from MC simulation and yields floated,
and take the difference of the fitted signal yields as
$\sigma_{\rm fit}$ for each $p^{*}_{\Xi^{-} \ell^{+}}/p^{*}_{\rm max}$ bin
($2.30\%-4.54\%$ for the electron mode and $2.34\%-5.10\%$ for the muon mode).
For $\Xi_{c}^{0}\to \Xi^{-} \pi^{+}$, we
estimate $\sigma_{\rm fit}$ by changing the range of the fit and the order of the background polynomial, and take the differences of the fitted signal yields as systematic uncertainties ($1.03\%-1.46\%$ depending on the $p^{*}_{\Xi^{-} \pi^+}/p^{*}_{\rm max}$ region).
By using the control sample $\Xi_c^{0} \to \Xi^{-} \pi^{+}$, the maximum difference in selection efficiency of the requirement $p^{*}_{\Xi^{-} \pi^{-}}/p^{*}_{\rm max}\textgreater0.45$ between weighted MC simulation based on $p^{*}_{\Xi^{-}X}/p^{*}_{\rm max}$ distribution from data and different signal MC simulations with different fragmentation functions in {\sc PYTHIA} generator~\cite{pythia} is 3.0\%, which is taken as the systematic uncertainty ($\sigma_{\varepsilon^{\rm pop}}$). For semileptonic decays, the uncertainties of the form factors in Ref.~\cite{LQCD} introduce a 3.1\% (3.6\%) uncertainty in the electron (muon) mode ($\sigma_{\rm FF}$). The change of the branching fraction measured with the sub-datasets with $p^{*}_{\Xi^{-}X}/p^{*}_{\rm max}\textgreater0.75$ that removes all background from $B$ decay is taken as the uncertainty associated with modeling of the $B$-decay background ($\sigma_{B\bar{B}}$) which is 2.5\% (6.3\%) for electron (muon) mode.
The systematic uncertainties $\sigma_{\rm PID}$ ($\sigma_{\rm fit}$) are added linearly (in quadrature) weighted by $\frac{N^{\Xi^{-}X}_{i}}{\varepsilon^{\Xi^{-}X}_{i}}$ and then
summed with $\sigma_{\varepsilon^{\rm pop}}$, $\sigma_{\rm FF}$, and $\sigma_{B\bar{B}}$ in quadrature to yield the total systematic uncertainty
($\sigma_{{\cal B}}$) for each $\Xi_{c}^{0}$ decay mode, which yields 4.6\%, 7.6\%, and 3.1\% for the electron, muon, and pion mode, respectively. The final systematic uncertainty on the branching fraction is the sum of the corresponding two $\sigma_{{\cal B}}$s in quadrature, which yields 5.6\% for ${\cal B}(\Xi_{c}^{0} \to \Xi^{-} e^{+} \nu_{e})$, and 8.2\% for ${\cal B}(\Xi_{c}^{0} \to \Xi^{-} \mu^{+} \nu_{\mu})$. The uncertainty of 28.9\% on ${\cal B}(\Xi_{c}^{0} \to \Xi^{-} \pi^{+})$~\cite{xic0_BR} is treated as an independent systematic uncertainty.
The total systematic uncertainty for ${\cal B}(\Xi_{c}^{0} \to \Xi^{-} e^{+} \nu_{e})/{\cal B}(\Xi_{c}^{0} \to \Xi^{-} \mu^{+} \nu_{\mu})$ is 6.8\% with the $\sigma_{B\bar{B}}$ negatively correlated.

The sources of systematic uncertainties in $\alpha^{\pm}$ include fitting procedures ($\sigma_{\rm fit}^{\alpha^{\pm}}$) and uncertainties on $\alpha_{\Xi^{\pm}}$ values ($\sigma_{\alpha_{\Xi^{\mp}}}^{\alpha^{\pm}}$).
$\sigma_{\rm fit}^{\alpha^{\pm}}$ are estimated to be 0.2\% with a toy MC method whose simulated distributions of $\alpha^{\pm}$ are found to be unbiased.
The uncertainties on $\alpha_{\Xi^{\pm}}$ values are $\sigma_{\alpha_{\Xi^{-}}}^{\alpha^{+}}$ = 2.1\% and $\sigma_{\alpha_{\bar{\Xi}^{+}}}^{\alpha^{-}}$ = 1.9\%~\cite{BES_alphaXi}, which are the leading systematic uncertainties. The final systematic uncertainties of $\alpha^{\pm}$ are $\sigma_{\alpha^{{\pm}}} = \sqrt{(\sigma_{\rm fit}^{\alpha^{\pm}})^{2} + (\sigma_{\alpha_{\Xi^{\mp}}}^{\alpha^{\pm}})^{2} }$. The systematic uncertainty $\Delta_{\AW_{CP}}$ is equal to $2\Delta r/(1-r)^{2}$. Here $r = \alpha^{+}/\alpha^{-}$, $ \Delta r = |r| \times \sqrt{\sigma_{\alpha^{+}}^{2} + \sigma_{\alpha^{-}}^{2} }$. Finally, the systematic uncertainties for $\alpha^{+},~\alpha^{-}$, and $\AW_{CP}$ are estimated to be 0.01, 0.01, and 0.014, respectively.

In summary, based on data samples of 89.5 and 711 ${\rm fb}^{-1}$ collected with the Belle detector at $\sqrt{s} = 10.52$ GeV and $\sqrt{s} = 10.58$, respectively, we measure the branching fractions of the $\Xi_{c}^{0} \to \Xi^{-} \ell^{+} \nu_{\ell}$ decays, $\Xi_{c}^{0}(\bar{\Xi}_{c}^{0}) \to \Xi \pi$ decay parameters $\alpha^{\pm}$, and the corresponding $CP$-asymmetry parameter $\AW_{CP}$. The measured branching fractions are ${\cal B}(\Xi_{c}^{0} \to \Xi^{-} e^{+} \nu_{e})=(1.31 \pm 0.04 \pm 0.07 \pm 0.38)\%$ and ${\cal B}(\Xi_{c}^{0} \to \Xi^{-} \mu^{+} \nu_{\mu})=(1.27 \pm 0.06 \pm 0.10 \pm 0.37)\%$. The ratio ${\cal B}(\Xi_{c}^{0} \to \Xi^{-} e^{+} \nu_{e})/{\cal B}(\Xi_{c}^{0} \to \Xi^{-} \mu^{+} \nu_{\mu})$ is $1.03 \pm 0.05 \pm 0.07$, which is consistent with the expectation of LFU~\cite{LQCD}. The measured $\Xi_{c}^{0}$ decay parameters are $\alpha^{+} = -0.64 \pm 0.05 \pm 0.01$ and $\alpha^{-} = 0.61 \pm 0.05 \pm 0.01$. The corresponding average absolute value of $\alpha^{\pm}$ is $0.63 \pm 0.03 \pm 0.01$ and the $CP$-asymmetry parameter $\AW_{CP}$ of $\Xi_{c}^{0} \to \Xi^{-} \pi^{+}$ decay is measured to be $0.024 \pm 0.052 \pm 0.014$. Here, the first and second uncertainties are statistical and systematic, respectively, while the third uncertainties on branching fractions are due to the uncertainty of ${\cal B}(\Xi_{c}^{0} \to \Xi^{-} \pip)$~\cite{xic0_BR}. The precision of the measurements of branching fractions of $\Xi_{c}^{0} \to \Xi^{-} \ell^{+} \nu_{\ell}$ and the $\alpha^{\pm}$ of $\Xi_{c}^{0} \to \Xi^{-} \pi^{+}$ is greatly improved compared to previous experimental results~\cite{arg_xic,clo_xic,clo_a_xic}. The measured $\AW_{CP}$ is consistent with no $CP$ violation. The semileptonic branching fraction ${\cal B}(\Xi_{c}^{0} \to \Xi^{-} \ell^{+} \nu_{\ell})$ is an important input used to constrain parameters of lattice QCD calculations~\cite{LQCD} and phenomenological models~\cite{xicBR_Theory1,xicBR_Theory2,xicBR_Theory3,xicBR_Theory4,xicBR_Theory5} of heavy-flavor baryon decays. As more precise measurements of ${\cal B}(\Xi_{c}^{0} \to \Xi^{-}\pi^{+})$ become available, the results presented in this Letter will allow the value of ${\cal B}(\Xi_{c}^{0} \to \Xi^{-} \ell^{+} \nu_{\ell})$ to be further improved.


Y. B. Li acknowledges the support from China Postdoctoral Science Foundation (2020TQ0079).
We thank the KEKB group for excellent operation of the
accelerator; the KEK cryogenics group for efficient solenoid
operations; and the KEK computer group, the NII, and
PNNL/EMSL for valuable computing and SINET5 network support.
We acknowledge support from MEXT, JSPS and Nagoya's TLPRC (Japan);
ARC (Australia); FWF (Austria);
the National Natural Science Foundation of China under
Contracts No. 11575017, No. 11761141009, No. 11975076, No. 12042509, No.12135005; the CAS Center for Excellence in Particle
Physics (CCEPP); MSMT (Czechia); CZF, DFG, EXC153, and VS (Germany);
DST (India); INFN (Italy);
MOE, MSIP, NRF, RSRI, FLRFAS project, GSDC of KISTI and KREONET/GLORIAD (Korea);
MNiSW and NCN (Poland); MSHE, Agreement 14.W03.31.0026 (Russia); University of Tabuk (Saudi Arabia); ARRS (Slovenia);
IKERBASQUE (Spain);
SNSF (Switzerland); MOE and MOST (Taiwan); and DOE and NSF (USA).


\begin{thebibliography}{99}
\bibitem{charge-conjugate} Inclusion of charge-conjugate states is implicit unless otherwise stated.
\end{thebibliography}
\end{document}